\documentclass[aps,prd,amsmath,floats,floatfix, twocolumn,
superscriptaddress,nofootinbib,showpacs]{revtex4-1}

\usepackage[T1]{fontenc}
\usepackage[utf8]{inputenc}
\usepackage{svg}
\usepackage{lmodern}
\usepackage{tensor}
\usepackage{verbatim}
\usepackage{siunitx}
\usepackage[caption=false]{subfig}
\usepackage{booktabs}
\usepackage{mathrsfs}
\usepackage{multirow}
\usepackage[dvipsnames, usenames]{xcolor}
\definecolor{linkcolor}{rgb}{0.0,0.3,0.5}
\usepackage[hypertexnames=false, unicode, colorlinks=true, linkcolor=linkcolor,
citecolor=linkcolor, filecolor=linkcolor,urlcolor=linkcolor,
pdfusetitle]{hyperref}
\usepackage[all]{hypcap}
\usepackage{graphicx}
\usepackage{xspace}
\usepackage{amssymb}
\usepackage[normalem]{ulem} 
\usepackage{bm} 
\usepackage{enumitem}
\usepackage{mathtools} 
\usepackage{txfonts}

\usepackage{microtype}
\usepackage{float}
\usepackage[english]{babel}
\usepackage{tikz}
\usetikzlibrary{decorations.pathmorphing}
\usetikzlibrary{decorations.markings}
\usepackage{orcidlink}

\begin{document}

\title{Relativistic tidal excitation of $g$- and $i$-modes in neutron star binaries}

\author{Binyu Zhong}
\email{Halley0627@sjtu.edu.cn}
\affiliation{Tsung-Dao Lee Institute, Shanghai Jiao-Tong University, Shanghai, 520 Shengrong Road, 201210, China}
\affiliation{School of Physics $\&$ Astronomy, Shanghai Jiao-Tong University, Shanghai, 800 Dongchuan Road, 200240, China}

\author{Zhiqiang Miao}
\email{zq_miao@sjtu.edu.cn}
\affiliation{Tsung-Dao Lee Institute, Shanghai Jiao-Tong University, Shanghai, 520 Shengrong Road, 201210, China}

\author{Zhen Pan}
\email{zhpan@sjtu.edu.cn}
\affiliation{Tsung-Dao Lee Institute, Shanghai Jiao-Tong University, Shanghai, 520 Shengrong Road, 201210, China}
\affiliation{School of Physics $\&$ Astronomy, Shanghai Jiao-Tong University, Shanghai, 800 Dongchuan Road, 200240, China}

\author{Huan Yang}
\affiliation{Department of Astronomy, Tsinghua University, Beijing 100084, China}

\date{\today}

\begin{abstract}
When the gravitational wave frequency of an inspiraling binary matches a normal-mode frequency of a neutron star (NS), the  mode can be resonantly excited. 
The excitation transfers orbital energy to the star and shifts the phase of the emitted gravitational-wave (GW) signal, thereby providing a potential probe of the NS interior. 
However, estimates of this phase shift in the literature vary by orders of magnitude due to different approximations and formulations used in calculating the tidal coupling strength. 
We revisit this problem using a fully relativistic scattering formalism, which serves as our benchmark calculation and provides a gauge-invariant mode response together with an effective tidal-mode coupling strength. We calculate the phase shifts induced by the excitation of stratification $g$-modes, crustal interface $i$-modes, and hadron-quark interface $i$-modes, and 
estimate their detectability with Fisher-matrix forecasts for current and next-generation detectors. 
In our models, direct $g$-mode excitation remains below the detection thresholds, even for the strongest stratification considered. Crust--core $i$-modes at large shear speeds provide the most favorable prospects and crust melting can produce a substantially larger phase shift. Hadron-quark $i$-modes remain below the next-generation detection thresholds.
\end{abstract}

\maketitle

\section{Introduction}
Binary neutron star (BNSs) are among the most important sources for gravitational wave (GW) detection.
The GW signal emitted during the inspiral phase encodes valuable information about the internal structure and equation of state (EoS) of neutron stars (NSs) through mutually induced tidal interactions~\cite{flanagan2008constraining,LIGOScientific:2017vwq,abbott2018gw170817}.
While quasi-static deformations dominate the tidal response during the early inspiral, dynamical tidal effects can become increasingly important as the binary approaches merger~\cite{Hinderer:2016eia}. In particular, when the tidal driving frequency matches a stellar oscillation frequency, resonant excitation can transfer orbital energy to the mode and leave a characteristic imprint on the GW phase~\cite{Lai:1993di,Reisenegger1994}.
Such resonances involving low-frequency modes have therefore been proposed as probes of  neutron star interior physics, such as possible hadron-quark phase transitions~\cite{Sotani:2001bb,Jaikumar:2021jbw,Miao:2023jqe,Counsell:2025hcv,Pereira:2025xsi,Lau:2020bfq,Constantinou:2021hba} and superfluidity~\cite{Yu:2016ltf,Yu:2017cxe}.

In Newtonian theory, resonant tidal excitation is well understood for a nonrotating star undergoing linear, adiabatic oscillations. 
The conservative perturbation problem is self-adjoint, so the stellar response can be expanded in an orthogonal set of normal 
modes~\cite{Chandrasekhar1964,Press1977,Lai:1993di}.
Each mode couples to the external tidal field through a tidal overlap integral $Q_a$ (Sec.~\ref{subsec: modesum});  together with the mode frequency and binary parameters, $Q_a$ determines the resonant phase shift. For a NS, however, reliable quantitative predictions require a consistent relativistic treatment of both the stellar oscillations and the tidal driving.

To this end, we employ the fully relativistic scattering formalism developed in Ref.~\cite{Miao:2025utd} to revisit the mode-excitation problem.
In the scattering formalism, the tidal interaction is reformulated as a wave-scattering problem, in which the external tidal field is represented by an ingoing wave and the stellar response is encoded in the outgoing component. 
This formulation provides a  gauge-invariant characterization of the tidal driving, with the response encoded in the analytic structure of the scattering coefficient.
It also enables a relativistic calculation of the tidal energy and defines an effective overlap integral $Q_a^{(\rm scatt)}$ through Eq.~(\ref{eq:scatterQ}). 
A causal relativistic theory of the stellar tidal response has recently been developed within the same framework~\cite{Miao:2026rie}.

For comparison, alternative evaluations of the tidal coupling have generally relied on approximate or reduced treatments. Hybrid calculations combine relativistic stellar backgrounds with Newtonian perturbations~\cite{Tsang:2011ad}, whereas other studies define generalized tidal overlap integrals within the relativistic Cowling approximation or related reduced frameworks~\cite{Counsell:2024pua,Counsell:2025hcv,Miao:2023jqe,Kuan:2021jmk,Pereira:2025xsi}.
Additional formulations include low-frequency expansions of the tidal coefficients~\cite{Pitre:2023xsr}, near-zone post-Newtonian boundary conditions~\cite{HegadeKR:2025qwj,HegadeKR:2026kku}, matching of the interior perturbation to the exterior tidal response~\cite{Andersson:2025iyd,Gao:2026sbm}, and descriptions of tidal scattering in the post-Minkowskian regime~\cite{Ivanov:2026icp,Ivanov:2024sds}.
In this work, the fully relativistic scattering calculation is used as the benchmark, and these alternative prescriptions are adopted only for comparison at fixed stellar models and mode families.

Here we apply the scattering formalism to low-frequency neutron star oscillations, focusing on $g$- and $i$-modes, and compare its predictions with those of the approaches discussed above. Given the uncertainties in the neutron star EoS, we adopt parameterized stellar models that capture the essential physics supporting these modes: composition stratification, crustal elasticity, and a density discontinuity (e.g., at a hadron-quark phase transition). These controlled models allow us to compare coupling prescriptions for matching models and to examine how the tidal coupling and resulting phase shift 
depend on stellar properties.

This paper is organized as follows. In Sec.~\ref{sec: Quasi-normal modes of neutron stars} we introduce the stellar models and their oscillation modes. We then 
describe the Newtonian mode-sum and relativistic scattering descriptions of the tidal coupling in Sec.~\ref{sec:formalisms}. We present the mode properties, coupling strengths, crust-melting estimates, and detectability forecasts in Sec.~\ref{sec:results}, and conclude in Sec.~\ref{sec:conclusion}. Appendix~\ref{app:perturbations} provides the linear perturbation equations for perfect-fluid, elastic-solid, and exterior vacuum regions. 
Appendix~\ref{app:andersson2025} 
summarizes the near-zone boundary-condition formulation used to obtain the comparison coupling $\sqrt{\mathcal{A}_n}$. 
Appendix~\ref{app:numerical} describes the boundary conditions, numerical integration, and quasi-normal-mode search.
Appendix~\ref{app:corner} illustrates the parameter correlations entering the Fisher forecast.
Throughout this paper, we use units with $c = G = 1$ and adopt the metric signature $(-,+,+,+)$.

\section{Quasi-normal modes of neutron stars}
\label{sec: Quasi-normal modes of neutron stars}
In this section, we first briefly review the relativistic perturbation of neutron stars. We then introduce parametrized neutron star models designed to study three classes of low-frequency oscillations: stratification-supported $g$-modes, characterized by the stratification strength $\delta$; crust--core (and also ocean--crust) interface ($i$-) modes, characterized by the crustal shear speed $v_s$; and hadron--quark $i$-modes, characterized by the energy-density jump $\Delta\varepsilon$. 

\subsection{Relativistic perturbation of neutron stars}
To calculate the oscillation modes, we consider linear polar perturbations of a static, spherically symmetric relativistic star. The background spacetime is described by
\begin{equation}
ds^2 = -e^\nu dt^2 + e^\lambda dr^2 + r^2 d\theta^2 + r^2 \sin^2\theta\, d\phi^2,
\end{equation}
and the stellar matter is modeled as a perfect fluid with energy-momentum tensor
\begin{equation}
T^{\rm{fluid}}_{ab} = (\varepsilon + p)u_a u_b + p g_{ab}\,,
\end{equation}
where $u_a$ is the fluid four-velocity, $p$ and $\varepsilon$ are the pressure and the energy density respectively.
The equilibrium configuration is then obtained by solving the Tolman--Oppenheimer--Volkoff equations, 
\begin{align}
    \frac{{\rm d} m}{{\rm d}r} &= 4\pi r^2\varepsilon\,,\\
    \frac{{\rm d}\nu}{{\rm d}r} &= \frac{m+4\pi r^3 p}{r(r-2m)}\,,\\
    \frac{{\rm d}p}{{\rm d}r} &= -(p+\varepsilon)\frac{\rm d\nu}{{\rm d}r}\,.
\end{align}
with $m(r)$ denotes the enclosed mass within $r$. The total gravitational mass of the neutron star is then $M =m(R)$, where $R$ is neutron star radius determined by $p(R) =0$.

In the Regge-Wheeler gauge~\cite{reggeStabilitySchwarzschildSingularity1957}, we write the metric perturbation as
\begin{equation}
\delta g_{ab}
=
-r^\ell
\begin{pmatrix}
H_0 e^\nu & i\omega r H_1 & 0 & 0 \\
i\omega r H_1 & H_2 e^\lambda & 0 & 0 \\
0 & 0 & r^2 K & 0 \\
0 & 0 & 0 & r^2 \sin^2\theta\, K
\end{pmatrix}
Y_{\ell m} e^{-i\omega t},
\end{equation}
where $H_0,H_1,H_2,K$ are functions of $r$. $Y_{\ell m}(\theta, \phi)$ are spherical harmonics. The fluid displacement vector can be written as
\begin{equation}
\boldsymbol \xi^i
=
r^\ell
\begin{pmatrix}
W r^{-1} e^{-\lambda/2} \\
- V r^{-2}\partial_\theta \\
- V (r^2 \sin^2\theta)^{-1}\partial_\phi
\end{pmatrix}
Y_{\ell m} e^{-i\omega t},
\end{equation}
where $W$, $V$ are also functions of $r$ and describe the radial and angular components of the displacement, respectively.
Substituting these expressions into the linearized Einstein equations,
\begin{equation}
\delta G_{ab}=8\pi\delta T^{\rm{fluid}}_{ab},
\end{equation}
yields the coupled perturbation equations~\cite{kips.thorneNONRADIALPULSATIONGENERALRELATIVISTIC,lindblomQuadrupolEoScillationsNeutron1983}, which are explicitly given in Appendix~\ref{app:perturbations} . We obtain the mode frequencies and eigenfunctions by solving these equations with the appropriate conditions at the stellar center, surface, and exterior boundary. The detailed numerical procedure is described in Appendix~\ref{app:numerical}.

\subsection{Stratification $g$-modes}
We model the background fluid star with an energy-polytrope EoS,
\begin{equation}
    p=K \varepsilon^\gamma,
    \label{eq:polytrope-EoS}
\end{equation}
where $K$ is the polytropic constant and $\gamma$ is the polytropic index.

To allow for buoyancy-supported oscillations, we introduce a non-barotropic perturbation by distinguishing the equilibrium and adiabatic indices,
\begin{align}
    \Gamma_0 &= \frac{p+\varepsilon}{\varepsilon}\left(\frac{\partial \ln p}{\partial \ln \varepsilon}\right)_0,\\
    \Gamma &= \frac{p+\varepsilon}{\varepsilon}\left(\frac{\partial \ln p}{\partial \ln \varepsilon}\right)_{\rm{ad}},
\end{align}
where $\Gamma_0$ is determined by the equilibrium structure and $\Gamma$ governs the response of fluid elements to adiabatic perturbations.
The restoring force is characterized by the relativistic Brunt-V\"ais\"al\"a frequency $N_{\rm BV}$,
\begin{align}
N_{\rm BV}^2 = e^{\nu-\lambda} \frac{p'}{\varepsilon + p} \, A\,,
\label{eq:Adef}
\end{align}
where the relativistic Schwarzschild discriminant is
\begin{equation}
    A = \frac{p'}{p} \left( \frac{1}{\Gamma_0} - \frac{1}{\Gamma} \right)\,.
\end{equation}
We quantify the stratification by the dimensionless parameter
\begin{equation}
\delta := \frac{\Gamma}{\Gamma_0}-1.
\end{equation}
We vary $\delta$ while keeping the equilibrium stellar model fixed.
A positive $\delta$ implies $\Gamma>\Gamma_0$ and $N_{\rm BV}^2>0$, corresponding to stable stratification that supports $g$-modes.
In this work we consider $5\times10^{-3}\leq\delta\leq10^{-1}$~\cite{Sun:2025zpj}.

\subsection{Crust-core and crust-ocean $i$-modes}
We consider a star with an elastic crust between a fluid core and a fluid ocean.
The equilibrium configuration is constructed with the same polytropic EoS as in the purely fluid case [Eq.~(\ref{eq:polytrope-EoS})].
Because the equilibrium star is strain-free, elasticity enters only at the perturbative level.
Unified-EoS calculations place the crust-core transition at energy densities of $67-81\,\rm{MeV/fm^3}$~\cite{Pearson:2018tkr}.
We locate the crust-core interface at $R_1$, defined by $\varepsilon(R_1)=75\,\rm{MeV/fm^3}$, and the crust-ocean interface at $R_2=0.99R$.

Within the crust, the perturbation of the stress-energy tensor includes an elastic contribution~\cite{5391529d-e323-3e10-8544-fab7336760a0,carterSpeedSoundHighPressure1973,Gao:2025aqo}:
\begin{equation}
\delta T_a^{\ b}
= \delta T_a^{\ b\,\rm{fluid}}
+ \delta \pi_a^{\ b}.
\end{equation}
For a Hookean solid, the elastic stress is related to the shear strain perturbation by
\begin{equation}
\delta\pi_a^{\ b}=-2\mu\delta s_a^{\ b},
\end{equation}
where $\mu$ is the shear modulus, $s_a^{\ b}$ is the trace-free shear strain tensor, and $\delta s_a^{\ b}$ is its Eulerian perturbation.
Following Refs.~\cite{krugerImpactRelativisticCowling2025,krugerSeismologyAdolescentNeutron2015,sotaniShearInterfaceModes2023,2015PhDT.......604K}, 
we express the elastic stresses in terms of radial functions $T_1$ and $T_2$,
\begin{equation}
T_1 Y_{\ell m} e^{-i\omega t} = \delta \pi_r^{\ r},
\qquad
\frac{1}{r} T_2 \partial_\theta Y_{\ell m} e^{-i\omega t}
= \delta \pi_r^{\ \theta}\,.
\end{equation}
The elastic stress couples these functions to the fluid and metric perturbations.
The resulting perturbation equations are given in Appendix~\ref{appsubsec: solideqs}, 
and the procedure used to match the solutions across the fluid--solid boundaries is described in Appendix~\ref{app:numerical}.

The shear modulus drops to zero at both the crust--core and crust--ocean boundaries. These interfaces support shear-dominated oscillation modes, which we refer to as crust--core and crust--ocean $i$-modes and denote by $i_1$ and $i_2$, respectively, below. To study how crustal elasticity affects these modes, we parametrize the shear modulus by the shear speed~\cite{carterSpeedSoundHighPressure1973},
\begin{equation}
v_s^2=\frac{\mu}{\varepsilon+p},
\end{equation}
and vary $v_s$ in our calculations. 
The shear speed is expected to lie in the range $10^{-3}\lesssim v_s\lesssim10^{-2}$~\cite{Strohmayer1991}.

\subsection{Hadron-quark $i$-modes}
We consider a hybrid star with a fluid quark core arising from a first-order phase transition. In this configuration, the pressure remains continuous across the hadron--quark interface, while the energy density has a finite jump. This discontinuity provides the buoyancy that supports a hadron--quark $i$-mode (also called the discontinuity $g$-mode in the literature~\cite{Miao:2023jqe,Pereira:2025xsi}).

We model the EoS of the quark core with a constant-speed-of-sound parameterization~\cite{PhysRevD.88.083013}, and the total EoS reads
\begin{equation}
    \varepsilon =
    \begin{cases}
 \left(p/K\right)^{1/\gamma} & p < p_t \\
\varepsilon_t+\Delta\varepsilon+c_s^{-2}(p-p_t) & p > p_t
    \end{cases}
\end{equation}
Here $p_t$ is the transition pressure, $\varepsilon_t=(p_t/K)^{1/\gamma}$ is the energy density on the hadronic side of the interface, and $\Delta\varepsilon$ is the jump to the quark phase. The parameter $c_s^2=dp/d\varepsilon$ is the squared sound speed in the quark phase, which we fix at $c_s^2=0.6$ throughout this work. Also we fix $\varepsilon_t=400~\rm{MeV/fm^3}$, and set $\Delta \varepsilon/\varepsilon_t=0.2-0.6$, meaning that $\Delta \varepsilon=80-240~\rm{MeV/fm^3}$, respectively.


\section{Tidal-mode coupling strength}
\label{sec:formalisms}
In this section, we briefly review the Newtonian mode-sum formalism for calculating tidal--mode coupling strengths and the relativistic extensions used in the literature. We then show how an effective coupling strength can be extracted from the scattering formalism.

\subsection{Newtonian Mode-sum Formalism}
\label{subsec: modesum}
In the Newtonian framework, the stellar response to an external tidal field can be described by expanding the Lagrangian displacement vector in terms of normal modes~\cite{Press1977,Lai:1993di},
\begin{equation}
\boldsymbol{\xi}(t,\boldsymbol{x}) = \sum_a q_a(t)\,\boldsymbol{\xi}_a(\boldsymbol{x}),
\label{eq:basis1}
\end{equation}
where $q_a(t)$ denotes the amplitude of a mode labeled by $a \equiv \{n,\ell,m\}$. 
The mode function $\boldsymbol{\xi}_a$ satisfies the eigenvalue problem
\begin{equation}
\mathcal{L}[\boldsymbol{\xi}_a] = -\omega_a^2 \rho\, \boldsymbol{\xi}_a,
\label{eq:basis2}
\end{equation}
where $\mathcal{L}$ is a self-adjoint operator and $\rho$ is the background density. 
The eigenfunctions form an orthonormal basis under the inner product
\begin{equation}
\int d^3x\, \rho\, \boldsymbol{\xi}_a^* \cdot \boldsymbol{\xi}_b = N\delta_{ab}\,,
\label{eq:basis3}
\end{equation}
where $N$ is a normalization constant. In this work, we adopt the convention $N=MR^\ell$
so that the overlap integral defined below is dimensionless.

Owing to this orthogonality, the fluid equation of motion can be projected onto a given mode, yielding
\begin{equation}
\ddot q_a + \omega_a^2 q_a = -\frac{1}{M R^\ell} \int d^3x\, \rho\, \boldsymbol{\xi}_a^* \cdot \nabla U_{\rm ext},
\label{eq: mode external couple}
\end{equation}
where $U_{\rm ext}$ denotes the external tidal potential. 
Upon decomposing $U_{\rm ext}$ in terms of tidal multipole moments $\mathcal{E}_{\ell m}$,
\begin{equation}
U_{\rm ext} = -\frac{1}{\ell!} r^\ell \mathcal{E}^{\ell m} Y_{\ell m},
\end{equation}
the equation of motion reduces to
\begin{equation}
\ddot q_a + \omega_a^2 q_a = - \frac{1}{\ell!}\mathcal{E}^{\ell m} Q_a,
\label{eq:Qdef1}
\end{equation}
where the overlap integral $Q_a$ is defined as
\begin{equation}
Q_a \equiv \frac{1}{M R^\ell} \int d^3x\, \rho\, \boldsymbol{\xi}_a^* \cdot \boldsymbol{\nabla} \left(r^\ell Y_{\ell m}\right).
\label{eq:Qdef2}
\end{equation}
The quantity $Q_a$ fully characterizes the coupling strength between the external tidal field and the mode $a$, and serves as the fundamental measure of mode excitation in the Newtonian formalism.

In the relativistic case, the mode-sum formalism is no longer strictly valid. 
This is because the oscillation modes of a compact star are generically quasi-normal modes, which include damping due to gravitational-wave emission. 
As a result, the perturbation problem is non-conservative, and the modes do not form a complete orthogonal basis in the usual sense.
Nevertheless, one may formally generalize the Newtonian definition of the overlap integral by retaining its projection structure. 
In this approach, the overlap integral is suggested to be defined as~\cite{Kuan:2021jmk,Counsell:2024pua}
\begin{equation}
Q_{a} = \frac{1}{ MR^\ell}\int d^3x \sqrt{-g}(\varepsilon + p)\xi_a^{i*} \nabla_i \left(r^\ell Y_{\ell m}\right)\,,
\label{eq:Qmodesum}
\end{equation}
with the normailzation 
\begin{equation}
    MR^\ell = \int d^3x\sqrt{-g}e^{-2\nu} (\varepsilon + p)\xi_a^{i*} \xi_{ai} \,.
\end{equation}

Although the expression for $Q_a$ given above is not generally justified for radiative modes of a neutron star, it can be used consistently within the relativistic Cowling approximation, where metric perturbations are neglected~\cite{Counsell:2024pua}. 
More recent studies show that imposing post-Newtonian boundary conditions in a buffer region renders the coupled fluid--metric perturbation problem self-adjoint, allowing relativistic tidal overlap integrals to be defined from the driven mode equations~\cite{HegadeKR:2025qwj,HegadeKR:2026kku}.

\subsection{Relativistic Scattering Formalism}
In contrast to the mode-sum approach, a fully relativistic scattering formalism has been developed to describe the transient excitation of stellar oscillations~\cite{Miao:2025utd}.
The calculation retains the coupled fluid and metric perturbations, and models the stellar response through waves propagating between past and future null infinity; it therefore does not rely on the Cowling approximation or on a hybrid treatment of the stellar background and perturbations.
In this framework, the tidal driving is modeled as a gravitational wave injected from past null infinity, which is scattered by the star and propagates to future null infinity. 
The response of the star can be described by the scattering coefficient $\mathcal{R}(\omega)$, defined as the ratio between outgoing and ingoing wave amplitudes at infinity,
\begin{equation}
    \mathcal{R}(\omega) = \frac{A_{\rm out}(\omega)}{A_{\rm in}(\omega)}\,.
\end{equation}
In particular, the stellar modes correspond to poles of $\mathcal{R}(\omega)$ in the complex frequency plane.
Near a mode frequency $\omega_a$, the scattering coefficient can be expanded as
\begin{equation}
\mathcal{R}(\omega) = \frac{\mathcal{W}_a}{\omega - \omega_a}+\text{regular part},
\end{equation}
where $\mathcal{W}_a$ denotes the residue of $\mathcal{R}$ at the pole. 
In general, the QNM frequency is a complex number, which we write the real and imaginary parts explicitly as $\omega_a= \omega_{aR}-i\gamma_a$.

In the scattering approach, the tidal energy associated with the resonant excitation of a given mode can be written as~\cite{Miao:2025utd}, 
\begin{equation}
    E_a = \frac{1}{32}\frac{(\ell+2)!}{(\ell-2)!}\frac{|\mathcal{W}_a|}{\dot \omega} |A_{\rm in}|^2\omega_{aR}^2\ .
\label{eq:E_scatt1}
\end{equation}
For comparison, in the mode-sum formalism the mode energy is given by
\begin{equation}
    E_a = \frac{\pi}{2}\frac{1}{\dot \omega}\frac{|Q_a|^2}{(\ell!)^2}|\mathcal{E}^{\ell m}|^2MR^{2\ell-2}\ ,
\label{eq:E_scatt2}
\end{equation}
By comparing the two, we can extract an effective tidal-mode coupling strength in the scattering formalism as
\begin{equation}
Q_a^{\rm (scatt)} \equiv \left[\frac{\left[(2\ell+1)!!\right]^2}{2\pi}\frac{\ell(\ell-1)}{(\ell+1)(\ell+2)}\frac{|\mathcal{W}_a|}{MR^{2\ell-2}\omega_{aR}^{2\ell}}\right]^{1/2}\ ,
\label{eq:scatterQ}
\end{equation}
where we have used the relation~\cite{Miao:2025utd}
\begin{equation}
    |\mathcal{E}^{\ell m}| = \frac{(\ell+1)(\ell+2)\ell!}{2(2\ell+1)!!}|A_{\rm in}|\omega_{aR}^{\ell+1}\ .
\label{eq:mathcalE}
\end{equation}
We note that for the fluid modes with weak damping considered in this work, where $\gamma_a \ll \omega_{aR}$, one finds that the residue satisfies $|\mathcal{W}_a|\approx 2\gamma_a$~\cite{Miao:2025utd}. 
In practice, we therefore replace $|\mathcal{W}_a|$ with $2\gamma_a$ in Eq.~(\ref{eq:scatterQ}).

\section{results}
\label{sec:results}

In this section, we study the modes introduced in Sec.~\ref{sec: Quasi-normal modes of neutron stars} by varying the parameters that govern their restoring forces: the stratification parameter $\delta$ for $g$-modes, the crustal shear speed $v_s$ for elastic $i$-modes, and the energy-density jump $\Delta\varepsilon$ for hadron-quark interface modes. We fix the stellar mass as $M=1.4M_{\odot}$ and restrict the radius of the models in $9-15~\rm{km}$. We calculate their tidal coupling strengths $Q_a$ using the fully relativistic scattering formalism [Eq.~(\ref{eq:scatterQ})] as our benchmark and compare the results with those obtained from the relativistic Cowling approximation [Eq.~(\ref{eq:Qmodesum})] and the surface-matching approach of Andersson et al.~\cite{Andersson:2025iyd}. Finally, we assess the resulting resonant phase shifts in the gravitational waveform.

\subsection{Mode Frequencies}

\begin{figure*}
    \includegraphics[width=\textwidth]{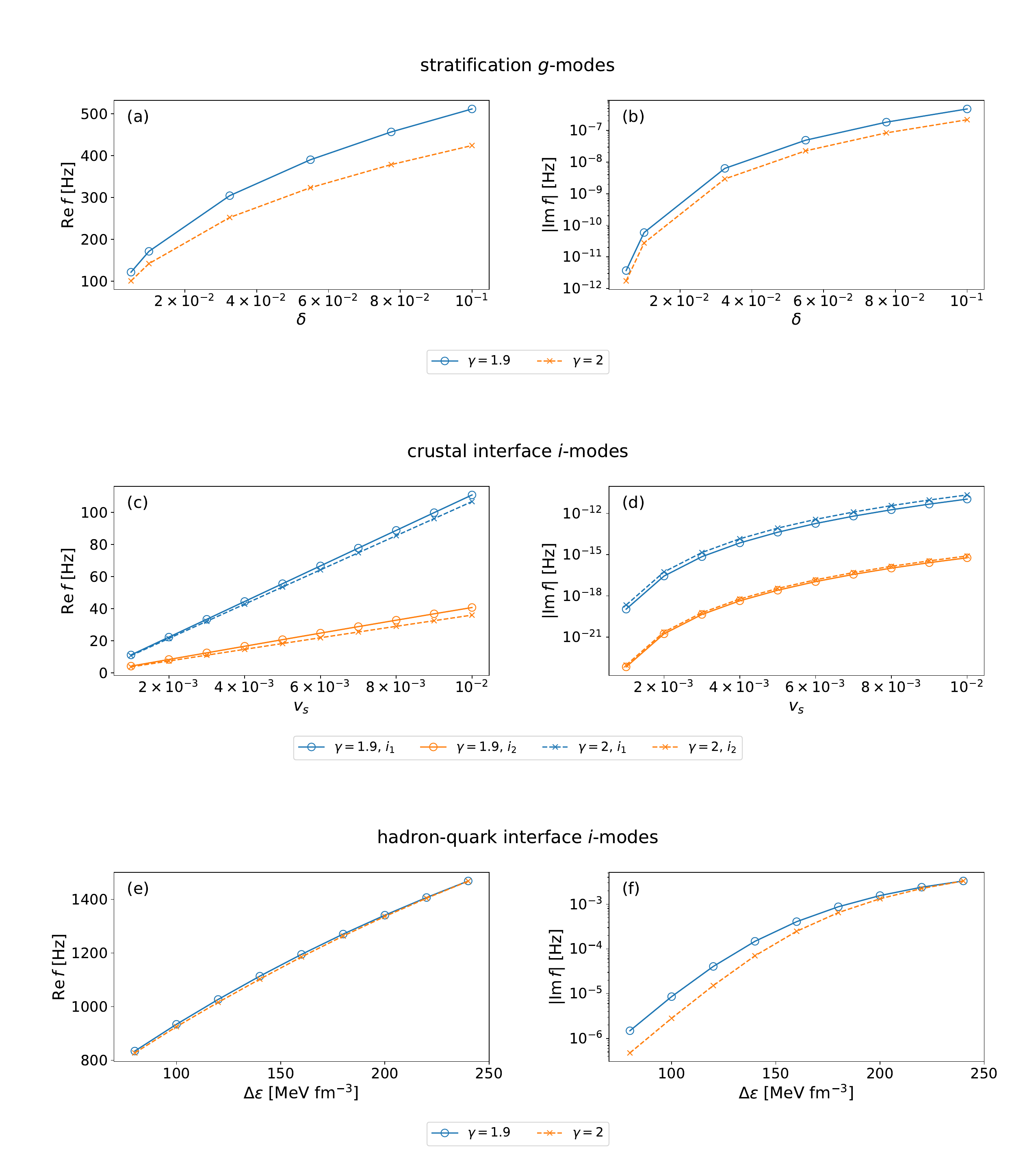}
    \caption{Mode frequencies of $g$- and $i$-modes in different neutron star models. (a/b): real/imaginary parts of stratification $g$-mode frequencies, (c/d): real/imaginary parts of crustal interface $i$-mode frequencies, (e/f): real/imaginary parts of hadron-quark interface $i$-mode frequencies.}
    \label{fig:qnm_f}
\end{figure*}

Figure~\ref{fig:qnm_f} shows how the complex mode frequencies vary with the parameters controlling each mode family. 
For each EoS family, we consider two polytropic models with $\gamma=1.9$ and $\gamma=2$, both with a stellar mass of $1.4M_\odot$ and radii of $13\,\rm{km}$ and $14\,\rm{km}$, respectively. We discuss the $g$-modes, $i_1$- and $i_2$-modes, and hadron--quark $i$-modes in turn.

For the $g$-modes of purely fluid stars, increasing the stratification parameter $\delta$ raises the real frequency, reflecting the stronger buoyant restoring force. More quantitatively, we find that the real frequency approximately follows
\begin{equation}
\operatorname{Re}f_g \propto \delta^{1/2}M^{1/2}R^{-3/2},
\end{equation}
consistent with the Newtonian scaling reported by Xu and Lai~\cite{xuResonantTidalExcitation2017}.
The magnitude of the imaginary frequency also increases with $\delta$, ranging from approximately $10^{-12}$ to $10^{-7}\,\rm{Hz}$ in the models considered. These small values indicate that the $g$-modes are weakly damped.

For stars with an elastic crust, the real frequencies of both $i_1$ and $i_2$ increase approximately linearly with the shear speed $v_s$. At a given $v_s$, the $i_2$ frequency is lower than that of $i_1$ by a factor of a few. The difference is much greater in their damping rates: $|\operatorname{Im}f_{i_2}|$ is several orders of magnitude smaller than $|\operatorname{Im}f_{i_1}|$, indicating that $i_2$ is much more weakly damped.

For stars with a hadron-quark interface, we continue to fix the stellar mass as $1.4M_{\odot}$; the stellar radius ranges from $12.6\,\rm km$ to $14.4\,\rm{km}$ over the range $\Delta \varepsilon$ considered. The two polytropic families yield nearly identical real mode frequencies when compared at the same $\varepsilon_t$, $\Delta\varepsilon$, and $c_s^2$, as in Fig.~\ref{fig:qnm_f}. This is because the hadron-quark interface-mode frequency is primarily controlled by the energy-density jump $\Delta\varepsilon$, with $\operatorname{Re}f_{i}$ scaling approximately as $(\Delta\varepsilon)^{1/2}$ when the other EoS parameters are held fixed. Varying $\varepsilon_t$ or $c_s^2$ changes the stellar structure and shifts the location of the phase transition, but has a subdominant effect on the mode frequency, as discussed by Miao et al.~\cite{Miao:2023jqe}.
The magnitude of the imaginary frequency is also much larger than for the crustal $i$-modes, ranging from $\sim 10^{-7}\,\rm Hz$ to $\sim 10^{-3}\,\rm Hz$ in our parameter range. As shown below, these modes also have larger excitation factors and tidal coupling strengths.

\subsection{Tidal-mode Coupling Strength}

\begin{figure*}
    \centering
    \includegraphics[width=\textwidth]{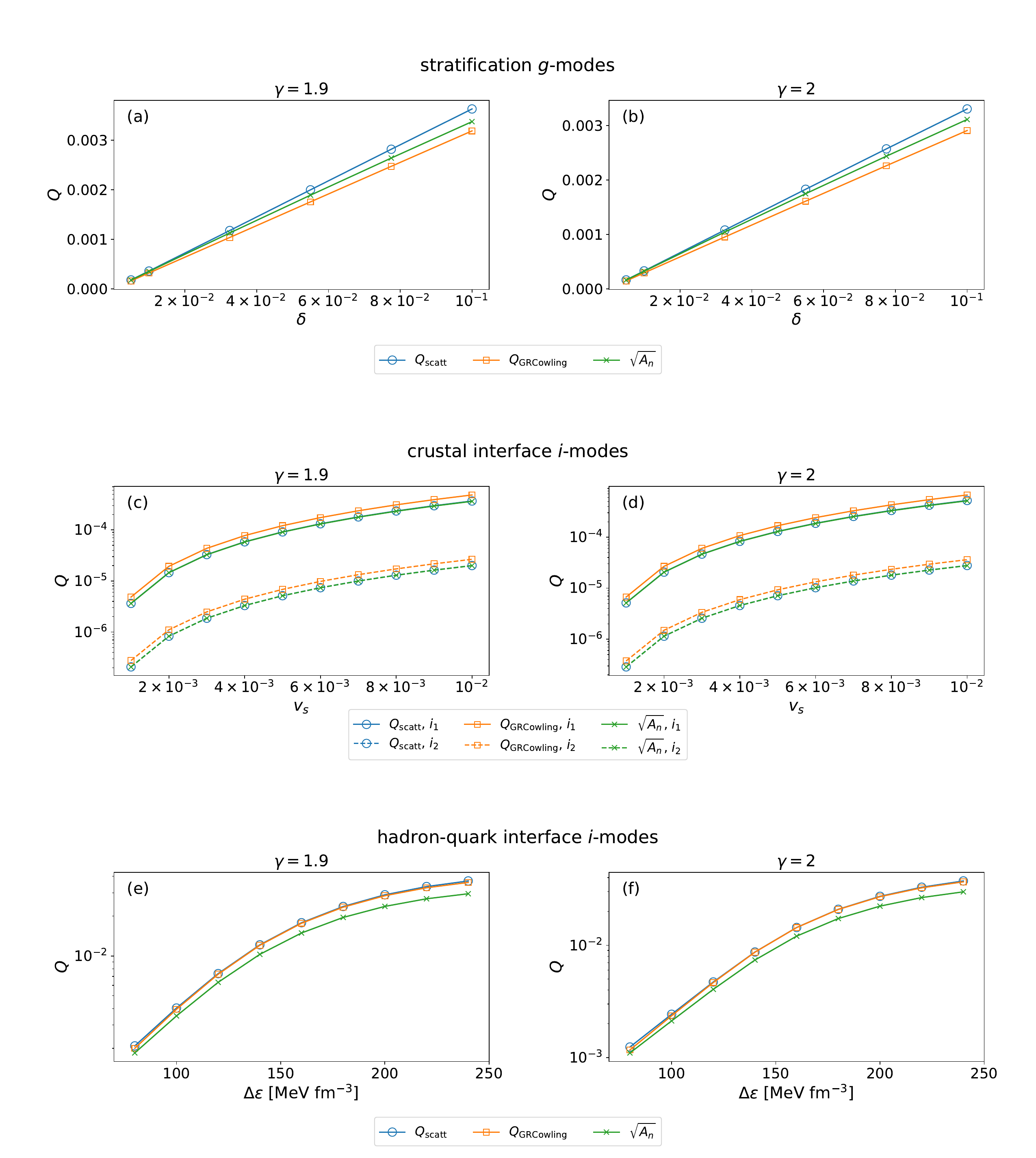}
    \caption{Comparison of several dimensionless tidal coupling strengths: the scattering coupling $Q_{\rm{scatt}}$, the GRCowling coupling $Q_{\rm{GRCowling}}$ and the near-zone coupling $\sqrt{\mathcal{A}_n}$. (a/b): coupling of the stratification $g$-modes, (c/d): crustal interface $i$-modes, (e/f): hadron-quark interface $i$-modes.}
    \label{fig:qnm_Q}
\end{figure*}

Here we compare the coupling strengths (normalized) in three different formalisms: the effective coupling strength $Q_{\rm{scatt}}$ defined in Eq.~(\ref{eq:scatterQ}), the relativistic Cowling generalization $Q_{\rm{GRCowling}}$ in Eq.~(\ref{eq:Qmodesum}) and the effective coupling strength $\sqrt{\mathcal{A}_n}$ proposed by Andersson et. al.~\cite{Andersson:2025iyd} (see Eq.~(\ref{eq:An}) in the Appendix~\ref{app:andersson2025}). Fig.~\ref{fig:qnm_Q} shows the results for the same stellar models shown in Fig.~\ref{fig:qnm_f}.
The three methods give similar trends with the stellar parameters and broadly agree on the relative strengths of the different modes. Taking the fully relativistic $Q_{\rm{scatt}}$ as the benchmark, the quantitative differences are small for low-frequency modes but become more pronounced at higher frequencies. In particular, they can reach a few tens of percent for the hadron--quark $i$-modes above $300~\rm{Hz}$.

We now turn to the coupling strengths of the individual mode families. 
For $g$-modes, the coupling increases from approximately $10^{-4}$ to a few times $10^{-3}$ as $\delta$ rises from $5\times10^{-3}$ to $10^{-1}$. These values are smaller than the Newtonian estimates of Xu and Lai~\cite{xuResonantTidalExcitation2017}, consistent with the relativistic reduction reported in Miao et al.~\cite{Miao:2025utd}. 
Among the crustal interface modes, $i_1$ couples one to two orders of magnitude more strongly than $i_2$. Both couplings increase with $v_s$, with that of $i_1$ reaching $\mathcal{O}(10^{-4})$ at $v_s=10^{-2}$.
The hadron--quark interface modes also have typical tidal-mode coupling strengths of order $10^{-3}$--$10^{-2}$, which increase with $\Delta\varepsilon$.

\subsection{Crust melting}

The tidal effect will progressively strengthen during the inspiral phase. This may cause the strain in the neutron star's solid layer to exceed a certain threshold at some point~\cite{Passamonti:2020fur}, leading to plastic deformation of the solid crust \cite{Chugunov2010}. This may opens up a new pathway for extracting energy from the orbit by melting the solid crust~\cite{Pan:2020tht}.

Here we use the framework introduced by Pan et al.~\cite{Pan:2020tht}. For crustal $i$-modes of our interest, we could evolve the forced oscillation equation~\cite{Lai:1993di}:
\begin{equation}
    \ddot{q}_{a}+\omega^2_a q_a=\frac{GM'}{R^3}Q_{a}\left(\frac{R}{D(t)}\right)^{\ell+1}W_{\ell m}e^{-im\Phi(t)},
    \label{eq:mode evolution}
\end{equation} 
where $q_a$ is the amplitude of the mode, $\omega_a=2\pi f_a$ is the angular frequency of the mode, $M'$ is the mass of the companion, $D(t),\Phi(t)=\int \Omega dt$ are the separation of the binary and the phase of the orbit respectively. The coefficients $W_{\ell m}$ are
\begin{equation}
\begin{aligned}
    W_{\ell m}
=
(-1)^{(\ell+m)/2}
\left[
\frac{4\pi}{2\ell+1}
(\ell+m)!(\ell-m)!
\right]^{1/2} \\
\times\left[
2^\ell
\left(\frac{\ell+m}{2}\right)!
\left(\frac{\ell-m}{2}\right)!
\right]^{-1}. 
\end{aligned} 
\end{equation}
The binary separation evolves, to leading order, by
\begin{equation}
    \dot{D}=-\frac{64}{5}\frac{MM'(M+M')}{D^3}
\end{equation}
and the orbital frequency $\Omega$ follows Kepler's law:
\begin{equation}
    \Omega=\left[\frac{(M+M')}{D^3}\right]^{1/2}.
\end{equation}
For each time $t$, we could get the strain by
\begin{equation}
    \sigma(t,r,\theta,\phi)=\sqrt{2\delta s_{ab}\delta s^{ab}} \ .
\end{equation}
We take the breaking strain to be $\sigma_b=0.1$~\cite{Chugunov2010,Horowitz:2009ya}. 
When $\sigma_{\max}$ exceeds this value, yielding begins and the dissipated energy 
heats the crust, leading to melting within a few orbital cycles~\cite{Pan:2020tht}. 
In analogy with resonant mode excitation, this process removes a fraction of 
the orbital energy and leaves a phase shift in the GW signal. 
The energy extracted from the orbit is estimated to 
be $E_{\rm melt}=\mathcal{O}(10^{47})\,\mathrm{erg}$~\cite{Pan:2020tht}.

 \begin{figure}
     \centering
     \includegraphics[width=\linewidth]{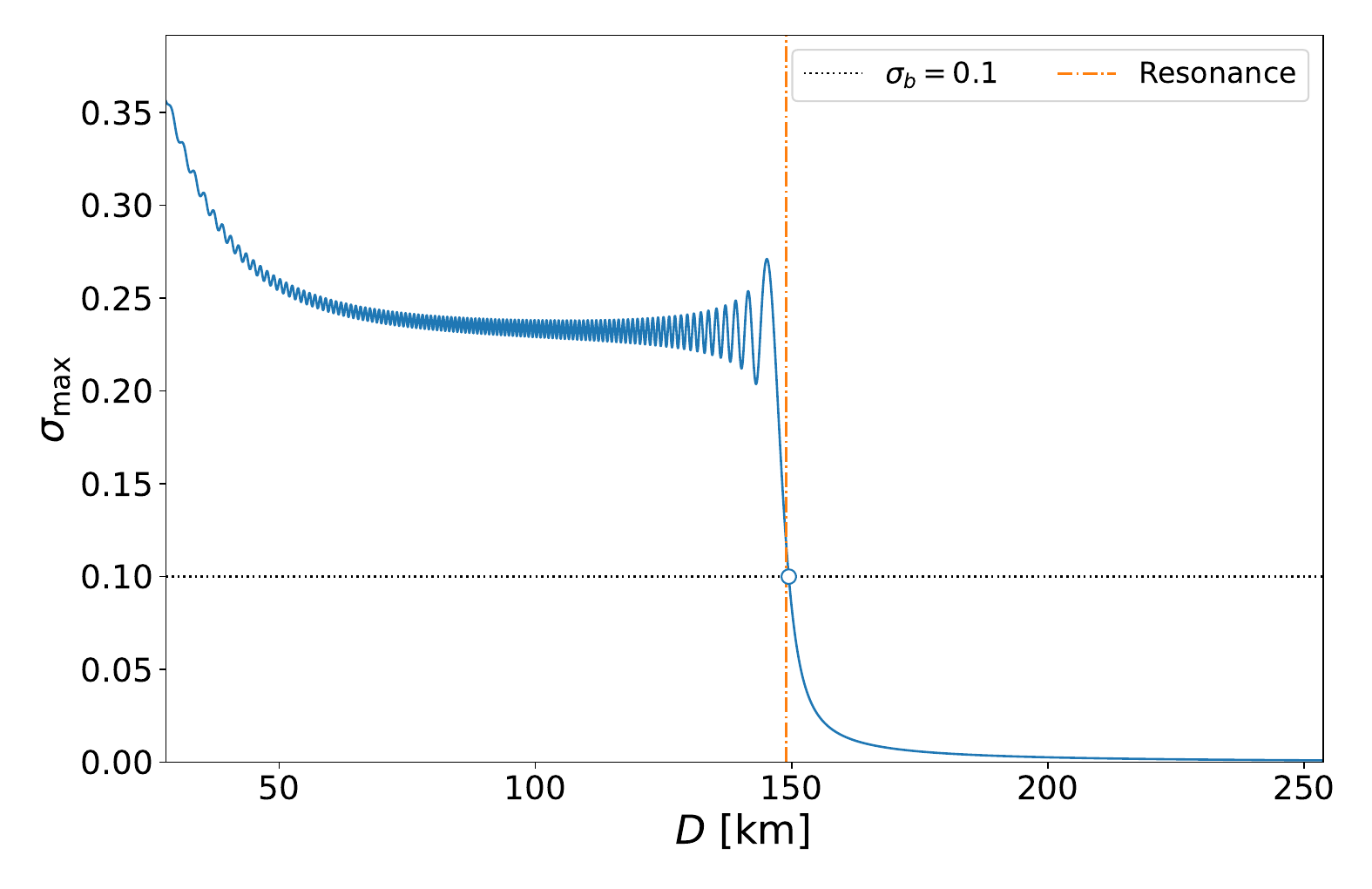}
     \caption{Maximum crustal strain $\sigma_{\max}$ as a function of binary separation $D$ for tidal excitation of the crust--core $i_1$-mode in an equal-mass binary with $M=M'=1.4M_\odot$. The stellar model has $\gamma=2$ and $v_s=0.01$. The horizontal dotted line marks the assumed breaking strain $\sigma_b=0.1$, the vertical dash-dotted line marks the resonance, and the open circle indicates where $\sigma_{\max}$ first reaches $\sigma_b$ as the binary inspirals. }
     \label{fig:melt}
 \end{figure}

\subsection{Phase Shifts in GWs and Detectability}
\begin{figure*}
    \centering
    \includegraphics[width=\textwidth]{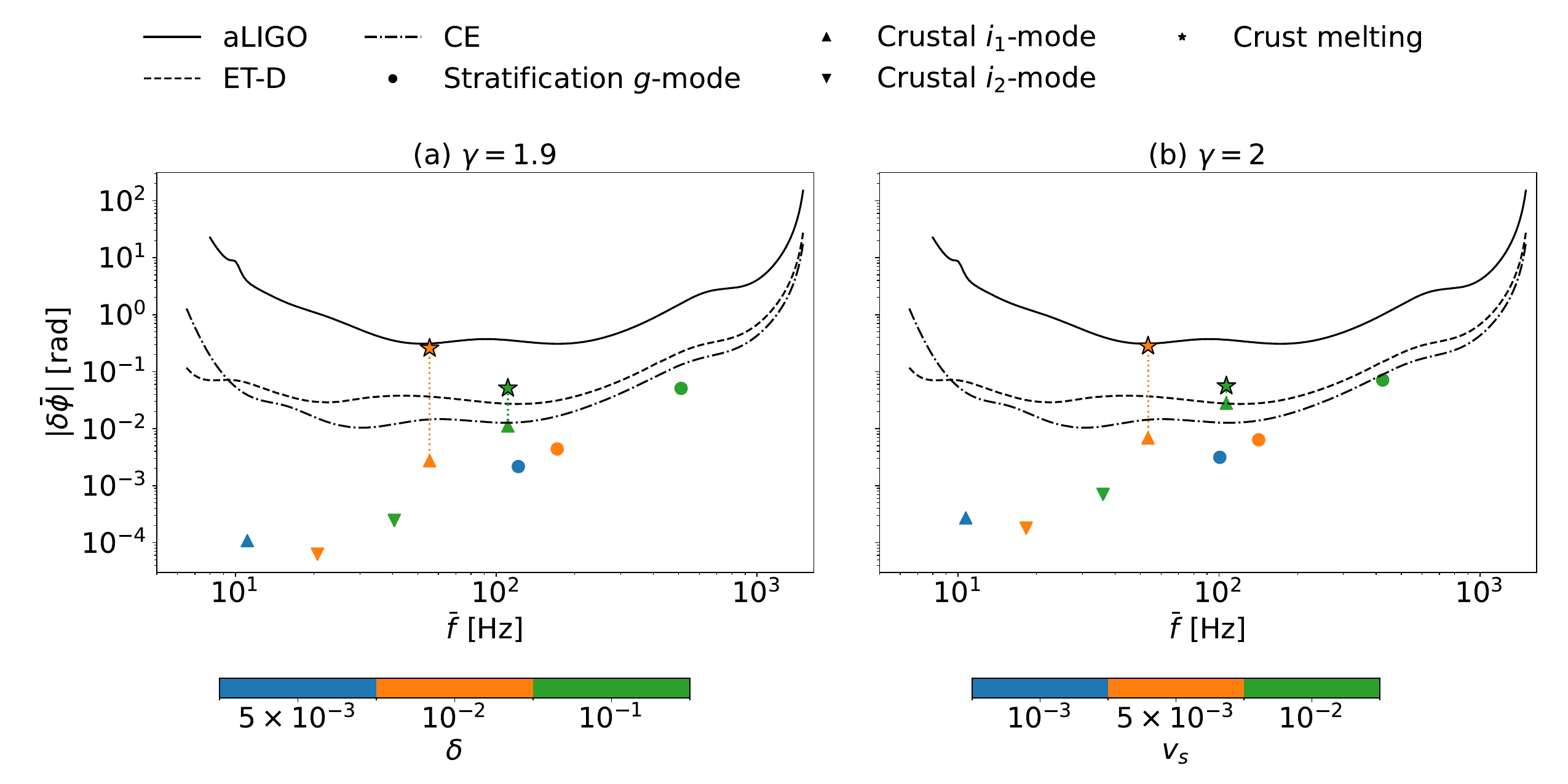}
    \caption{Total phase shifts of stratification $g$-modes and crustal interface $i$-modes. Modes with frequencies below $5\,\rm{Hz}$ are not shown. A point is potentially detectable when it lies above the curve for a given detector. Star markers show phase shifts produced when tidal excitation causes crust melting. Panels show the (a) $\gamma=1.9$ and (b) $\gamma=2$ stellar families.  }
    \label{fig:delta_phi}
\end{figure*}

\begin{figure*}
    \label{fig:delta_phi2}
    \centering
    \includegraphics[width=\textwidth]{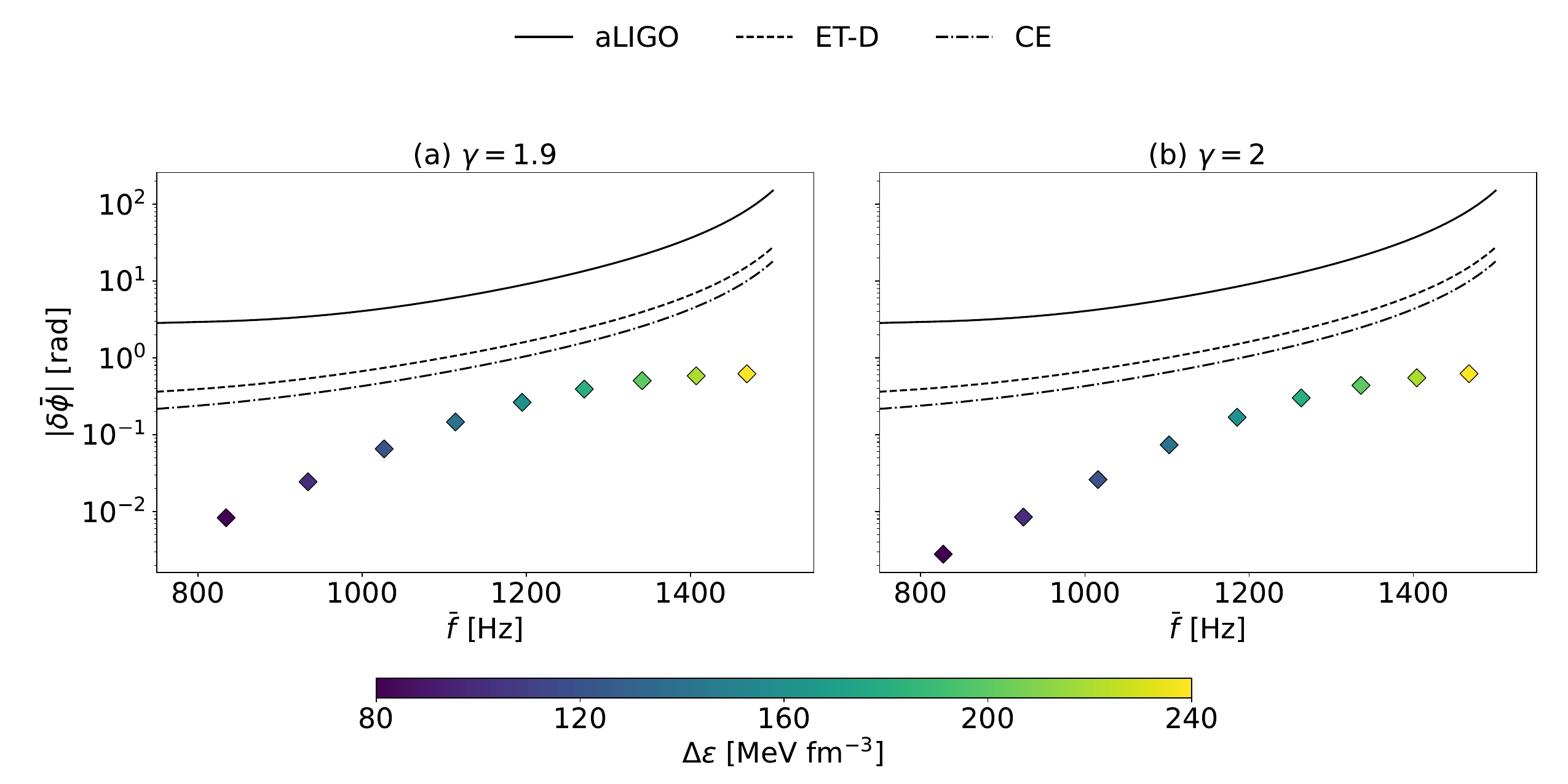}
    \caption{Same as Fig.~\ref{fig:delta_phi}, but for the hadron-quark $i-$modes.}
\end{figure*}

Resonant excitation transfers an energy $\Delta E_a>0$ from the orbit to mode $a$, thereby accelerating the inspiral. The resulting GW phase shift due to mode excitation in one star is
\begin{equation}
    \delta\phi_a
    =-\frac{2\pi f_a\Delta E_a}{P_{\rm{GW}}(f_a)}\,,
    \label{eq:delta_phi}
\end{equation}
where $f_a$ is the GW frequency at resonance.
At leading quadrupole order, the GW radiation power of a quasi-circular binary is~\cite{PhysRev.131.435}
\begin{equation}
    P_{\rm{GW}}(f_a)
    =\frac{32}{5}\left(\pi\mathcal{M}f_a\right)^{10/3},
    \label{eq:power}
\end{equation}
with chirp mass $\mathcal{M}=(MM')^{3/5}(M+M')^{-1/5}$.
Substituting the mode energy from Eq.~(\ref{eq:E_scatt2}) into Eq.~(\ref{eq:delta_phi}) yields
\begin{equation}
\begin{aligned}
    \delta\phi_a
    &=-\frac{5\pi^2}{2048}Q_a^2\omega_a^{-2}\frac{R^2}{M^4}\frac{2q}{1+q} \\
    &=-0.043 \left(\frac{2q}{1+q}\right)
\left(\frac{Q_a}{10^{-3}}\right)^2
\left(\frac{f_a}{100\,\rm{Hz}}\right)^{-2}
M_{1.4}^{-4}
R_{12}^2\,,
\end{aligned}
\label{eq:total_delta_phi}
\end{equation}
where $q=M'/M$, $M_{1.4}=M/1.4M_\odot$, and $R_{12}=R/12\,\rm{km}$.

For a crustal interface mode, crust melting can change the energy extracted from the orbit. If melting extracts an energy $E_{\rm{melt}}$ at GW frequency $f_{\rm{melt}}$, the associated phase shift for one star is~\cite{Pan:2020tht}
\begin{equation}
\begin{aligned}
\delta\phi_{\rm{melt}}
&=
-\frac{2\pi f_{\rm{melt}}E_{\rm{melt}}}
{P_{\rm{GW}}(f_{\rm{melt}})}\\
&=
-\frac{0.034}{q^2}
\left(\frac{1+q}{2}\right)^{2/3}\left(\frac{f_{\rm{melt}}}{100\,\rm{Hz}}\right)^{-7/3}
E_{47}M_{1.4}^{-10/3},
\end{aligned}
\end{equation}
where $E_{47}=E_{\rm{melt}}/(10^{47}\,\rm{erg})$.

Accounting for the excitation in both stars, the total phase correction to the binary GW waveform $h(f)=\mathcal{A}e^{i\Phi(f)}$ can be written as~\cite{Pan:2020tht,Yu:2017cxe}
\begin{equation}\label{eq:phase-corr}
    \begin{aligned}
        \delta \Phi(f) &=  \sum_{i=1,2}\delta\phi_i \left(1-\frac{f}{f_i}\right)\Theta(f-f_i) \\
        & \approx \delta\bar{\phi} \left(1-\frac{f}{\bar{f}}\right)\Theta(f-\bar{f}), \\ 
    \end{aligned}
\end{equation}
where $\Theta$ is the Heaviside step function, $\delta\phi_i$ is the phase shift produced by excitation of the relevant mode in star $i$, and $f_i$ is the corresponding resonance frequency.
The total phase shift and effective resonance frequency are defined by
\begin{equation}
    \delta\bar{\phi}=\sum_{i}\delta\phi_i\,,
    \qquad
    \frac{\delta\bar{\phi}}{\bar{f}}=\sum_i\frac{\delta\phi_i}{f_i}\,.
\end{equation}
Below each resonance, $\delta\Phi_i=0$. Above it, the correction separates into a constant phase offset and a term linear in frequency.
These terms are corrections to the coalescence phase $\phi_c$ and time $t_c$, so the post-resonance phasing can be absorbed into the redefined quantities
\begin{equation}
    \tilde{\phi}_c = \phi_c+\delta\bar\phi\,,
    \qquad
    \tilde{t}_c = t_c + \frac{\delta\bar\phi}{2\pi\bar f}\,.
\end{equation}
The post-resonance phase therefore does not determine $\delta\bar\phi$ by itself: one must compare the phasing before and after resonance.
This observation motivates the Fisher analysis below, in which $\phi_c$, $t_c$ and other waveform model parameters are marginalized over together with the resonance parameters.

Following Refs.~\cite{Yu:2017cxe,cutler1994gravitational,Miao:2023jqe}, we compute the Fisher matrix for a waveform $h$ described by parameters $\{\theta_i\}$:
\begin{equation}
    \Gamma_{ij}
    =\left(\frac{\partial h}{\partial\theta_i}
    \bigg|\frac{\partial h}{\partial\theta_j}\right),
\end{equation}
where the noise-weighted inner product is
\begin{equation}
    (a|b)=4\,\mathrm{Re}\int
    \frac{\tilde a^{*}(f)\tilde b(f)}{S_n(f)}\,df.
\end{equation}
Here $S_n(f)$ is the detector's one-sided noise power spectral density. Treating $\delta\bar{\phi}$ and $\bar f$ as parameters of the resonant correction, we obtain the marginalized uncertainty
\begin{equation}
    \sigma_{\delta\bar{\phi}}=\sqrt{(\Gamma^{-1})_{\delta\bar{\phi}\delta\bar{\phi}}}.
    \label{eq:1sigma}
\end{equation}
We regard the resonance-induced phase shift as potentially detectable when $|\delta\bar{\phi}|\gtrsim\sigma_{\delta\bar{\phi}}$.

In this work, we consider a nonspinning, face-on $1.4M_\odot$-$1.4M_\odot$ binary at a luminosity distance of $D_L=100 \ \rm{Mpc}$, with tidal deformabilities fixed at $\Lambda=\Lambda'=400$. 
For each injected value of $\bar f$, we use the waveform parameters
\begin{equation}
    \mathbf{\Theta}
    =\{\mathcal M,\eta, \tilde{\Lambda},t_c,\phi_c,D_L,\delta\bar{\phi},\bar f\}\,,
\end{equation}
where $\eta=MM'/(M+M')^2$ is the symmetric mass ratio and 
\begin{equation}
    \tilde\Lambda = \frac{16}{13}\frac{(1+12q)\Lambda +(12+q)q^4\Lambda'}{(1+q)^5}\,,
\end{equation}
is the mass-weighted tidal deformability.
The baseline waveform is \texttt{IMRPhenomD\_NRTidal}~\cite{PhysRevD.96.121501}, augmented by the resonant correction in Eq.~(\ref{eq:phase-corr}).
We assume an optimal antenna response ($F_+=1$, $F_\times=0$) and integrate over $f_{\min}=5\,\rm{Hz}$ to $f_{\rm ISCO}\simeq 4400\ {\rm Hz}/[(M+M')/M_\odot ]\approx1571\,\rm{Hz}$~\cite{cutler1994gravitational}.
For the design sensitivities of aLIGO, the Einstein Telescope (ET-D), and Cosmic Explorer (CE)~\cite{LIGOScientific:2014pky,Hild:2010id,LIGOScientific:2016wof}, the corresponding optimal signal-to-noise ratios (SNRs) are 49, 502, and 1495, respectively.

Figure~\ref{fig:delta_phi} shows the marginalized Fisher uncertainty $\sigma_{\delta\bar{\phi}}$ as a function of the effective resonance frequency $\bar f$ for each detector.
To assess how much this threshold is increased by parameter correlations, we compare it with an idealized split-SNR estimate,
\begin{equation}
    \sigma_{\rm split}
    =\sqrt{\rho_{<}^{-2}+\rho_{>}^{-2}}\,,
\end{equation}
where
\begin{equation}
\begin{aligned}
\rho_{<}^{2}
&= 4 \int_{f_{\min}}^{\bar f}
   \frac{|\tilde{h}(f)|^{2}}{S_n(f)} \, \mathrm{d}f, \\
\rho_{>}^{2}
&= 4 \int_{\bar f}^{f_{\rm ISCO}}
   \frac{|\tilde{h}(f)|^{2}}{S_n(f)} \, \mathrm{d}f.
\end{aligned}
\label{eq:split_snr}
\end{equation}
and $\tilde{h}(f)$ is the frequency-domain GW strain.
This estimate represents the information available from separately comparing the waveforms below and above resonance when all other parameters are known; it neglects the joint fit of the phase, timing, and intrinsic parameters.
In our models, the full Fisher uncertainty is typically about an order of magnitude larger than $\sigma_{\rm split}$ over the displayed frequency range.
The gap reflects the combined effect of parameter degeneracies.
The phase and timing offsets produced above resonance can be partly absorbed into $\phi_c$ and $t_c$, while the corner plot in Fig.~\ref{fig:full parameter corner} shows that $\delta\bar{\phi}$ is also strongly correlated with the effective tidal deformability $\tilde{\Lambda}$ and the resonance frequency $\bar f$.

Figures~\ref{fig:delta_phi} and~\ref{fig:delta_phi2} compare the predicted phase shifts and resonance frequencies with the Fisher thresholds.
In our models, no stratification $g$-mode or crust-ocean $i_2$-mode excitation reaches the next-generation thresholds.
The direct crust-core $i_1$-mode signal also remains below the ET-D threshold in the $\gamma=1.9$ family; in the $\gamma=2$ family, only the largest-shear model reaches approximately the ET-D threshold and lies above the CE threshold.
Crust melting changes this conclusion for the two largest shear speeds: the associated phase shifts are approximately $0.05$-$0.3\,\rm{rad}$, placing them above the ET-D and CE thresholds, although still below the aLIGO threshold in these examples.
For $g$-modes, increasing the stratification raises the predicted phase shift, but even the strongest cases with $\delta = 0.1$ remain below the CE threshold.
Despite their relatively large tidal coupling, hadron-quark $i$-modes are found with high frequencies and remain well below all three thresholds in our analysis.

These Fisher thresholds are broadly consistent with those of Gittins et al.~\cite{Gittins:2026ntx}, who found that ET-D could identify resonances in the $5$--$300\,\rm{Hz}$ range, with sensitivity to phase shifts as small as $\sim0.03\,\rm{rad}$ in  favorable events. For comparison, our marginalized ET-D and CE thresholds are approximately $3\times10^{-2}$ and $10^{-2}\,\rm{rad}$ at $100\,\rm{Hz}$, 
but increase to about $0.2$ and $0.1\,\rm{rad}$ at $500\,\rm{Hz}$.
However, our calculations do not support the suggestion that the $g$-modes considered here could produce detectable signals.~\footnote{An illustrative estimate in Ref.~\cite{Gittins:2026ntx}  
combines $f_a\sim100\,\rm{Hz}$ with $Q_a\sim10^{-5}$-$10^{-3}$ ~\cite{Counsell:2024pua} and can yield $\delta\bar\phi\sim0.08$ when the upper end of the coupling range is used. 
In the present mode calculations, however, $Q_a$ and $f_a$ are not independent: a coupling near $10^{-3}$ occurs at several hundred hertz rather 
than at $100\,\rm{Hz}$. Accounting for this relation and the scaling $\delta\phi_a\propto Q_a^2/f_a^2$, the predicted phase shifts remain below the 
illustrative $\sim0.03$ detection threshold used in that comparison.}
With parameters $\delta\bar\phi$ and $\bar f$ obtained from the mode calculations, none of the $g$-mode models considered here reaches the next-generation thresholds, even at the strongest stratification.
Our prospects for hadron-quark $i$-modes are less optimistic than those reported by Counsell et al.~\cite{Counsell:2025hcv}.
Part of this difference reflects the sensitivity criterion: their estimate uses a pointwise characteristic strain for a source at $40\,\rm{Mpc}$, whereas our marginalized Fisher analysis accounts for correlations between $\delta\bar{\phi}$ and the effective tidal deformability, coalescence phase and time, and resonance frequency.
At the same source distance, their threshold near $1\,\rm{kHz}$ is approximately an order of magnitude lower than our Fisher estimate.

\section{Discussion and conclusion}
\label{sec:conclusion}

In this work, we have revisited the resonant tidal excitation of low-frequency neutron-star modes in coalescing compact binaries. Using the scattering formalism, we calculated the mode energy and an effective tidal-mode coupling strength within a fully relativistic treatment of the stellar response. We considered stratification-supported $g$-modes, crustal $i$-modes, and hadron--quark $i$-modes, varying their key physical parameters in a set of parameterized stellar models. 
We then assessed the detectability of the resulting GW phase shifts with current and next-generation detectors.
Among the modes considered, the crust-core $i_1$-mode is the most promising target for next-generation detectors. Its direct phase shift reaches the ET-D threshold only in the largest-shear, $\gamma=2$ model, while crust melting can raise the signal above the ET-D and CE thresholds. The prospects for detecting stratification-supported $g$-modes and hadron-quark $i$-modes remain less favorable.

There could be several improvements to the present work. First, more realistic stellar models could refine the quantitative predictions by capturing the interplay among stratification, crustal elasticity, and a density discontinuity, which our parameterized models treat separately. Their interplay may qualitatively change the mode spectrum. For example, Gao et al.~\cite{Gao:2025aqo} found that the crust--core $i_1$ mode disappears in the stratified models they studied and is replaced by gravity modes with mixed gravity--interface character. 
This motivates microphysically informed models that determine the composition gradients, crustal shear modulus, and crust--core transition consistently~\cite{Sun:2025zpj}. 
Superfluidity may also suppress $g$-modes supported by the usual proton-fraction gradient~\cite{Lee1995,Andersson:2001bz,Prix:2002fk}, although muon--electron stratification can support a distinct population of $g$-modes with higher frequencies~\cite{Kantor:2014lja,Yu:2016ltf}. Their tidal excitation warrants a separate study.

Second, we currently calculate the crustal strain using a Newtonian mode-amplitude evolution equation [Eq.~(\ref{eq:mode evolution})], while obtaining the tidal-mode coupling strength from a relativistic scattering calculation. A fully consistent treatment would require a relativistic evolution equation. The recently developed causal relativistic theory of tidally driven mode evolution~\cite{Miao:2026rie} offers a possible route, provided it is extended to elastic matter. This would enable a more reliable assessment of crust failure and its associated GW phase shift.

\begin{acknowledgments}
We thanks professor Christian J. Kr\"uger and Hajime Sotani for the useful discussion. We acknowledge DeepSeek v4.1 Flash for assistance with the numerical analysis. All results are verified by the authors. We use Python with NUMPY\cite{Harris_2020}, SCIPY\cite{Virtanen_2020} and MATPLOTLIB\cite{4160265} for numerical calculations and plotting. ZM is supported by the Postdoctoral Innovation Talent Support Program of CPSF (No. BX20240223) and CPSF funded project (No. 2024M761948). ZP is supported by the National Natural Science Foundation of China (No. 12673052).
\end{acknowledgments}

\onecolumngrid
\appendix

\section{Perturbation Equations}\label{app:perturbations} 

\subsection{Perturbation Equations for Perfect Fluid Regions}
We could get the equations from the linearized Einstein equations. We use the form of the equations by Lindblom and Detweiler \cite{lindblomQuadrupolEoScillationsNeutron1983,detweilerNonradialPulsationsGeneral1985}. The perturbation equations has 4 ODEs:

\begin{equation}
H_1' =
\left[\frac{1}{2}(\lambda' - \nu') - \frac{\ell+1}{r}\right] H_1
+ \frac{e^\lambda}{r} \left[ H_0 + K - 16\pi(\varepsilon + p)V \right]
\end{equation}

\begin{equation}
K' =
\frac{1}{r} H_0 + \frac{n+1}{r} H_1
+ \left(\frac{1}{2}\nu' - \frac{\ell+1}{r}\right) K
- \frac{8\pi(\varepsilon + p)}{r} e^{\lambda/2} W
\end{equation}

\begin{equation}
\begin{aligned}
W' &=
-\left(\frac{\ell+1}{r} + \frac{p'}{\Gamma p}\right) W+ r e^{\lambda/2}
\Biggl[
\frac{\varepsilon + p}{\Gamma p}
\left(e^{-\nu}\omega^2 V + \frac{1}{2} H_0\right)
- \frac{2(n+1)}{r^2} V
 + \frac{1}{2} H_0 + K
\Biggr]
\end{aligned}
\end{equation}

\begin{equation}
\begin{aligned}
X' &=
-\frac{\ell}{r} X
+ \frac{1}{2}(\varepsilon + p)e^{\nu/2}
\left[
\left(\frac{1}{r} - \frac{\nu'}{2}\right) H_0
+ r\omega^2 e^{-\nu} V
+ \frac{n+1}{r} H_1
\right]
+ \left[
\frac{3}{2}\nu' - \frac{1}{r}
\right] K
- \nu' \frac{\ell(\ell+1)}{r^2} V \\
&\quad
- \frac{1}{r} \left[
8\pi(\varepsilon + p)e^{\lambda/2}
+ 2\omega^2 e^{\lambda/2-\nu}
- r^2\left(r^{-2}e^{-\lambda/2}\nu'\right)'
\right] W
\end{aligned}
\end{equation}

Where the prime denotes that the derivative to $r$. $n=\frac{1}{2}(\ell-1)(\ell+2)$.
We also have 3 algebraic equations:

\begin{equation}
\omega^2(\rho + p)V
=
e^{\nu/2} X
- \frac{1}{r} p' e^{-\nu/2} W
- \frac{1}{2}(\rho + p)e^\nu H_0
\end{equation}
\begin{equation}
\begin{aligned}
\Bigl[
(n + 1)r - \frac{r e^{-\lambda}}{2}(r\lambda' + 2)
\Bigr] H_0
&=
r^2 e^{-\lambda}
\Bigl(
\omega^2 r e^{-\nu}
- \frac{n + 1}{2}\nu'
\Bigr) H_1
+ [n r 
- \omega^2 r^3 e^{-\nu} 
- \frac{1}{4} r^2 e^{-\lambda}\nu'(r\nu' - 2)] K \\
&\quad
+ 4\pi r^2 (\varepsilon + p)e^{-\lambda/2} W
+ 8\pi r^3 (\varepsilon + p)\omega^2 e^{-\nu} V
\end{aligned}
\end{equation}
\begin{equation}
H_2 = H_0
\end{equation}

As discussed in \cite{Gao:2025aqo,2015PhDT.......604K}, we could use $V$ instead of $X$ as the variable in order to avoid the numerical problem in low-frequency domain. The corresponding equation is:

\begin{equation}
V' =
\left(-A + \nu' - \frac{\ell}{r}\right) V
- \frac{A e^\nu}{2\omega^2}
\left[
H_0 + \frac{\nu'}{r e^{\lambda/2}} W
\right]
+ r H_1
- \frac{e^{\lambda/2}}{r} W
\end{equation}   

$A$ is defined in Eq. (\ref{eq:Adef}).
\subsection{Perturbation Equations for Elastic Solid Crust Regions}
\label{appsubsec: solideqs}
Here we refer \cite{krugerImpactRelativisticCowling2025,krugerSeismologyAdolescentNeutron2015,Gao:2025aqo,2015PhDT.......604K} and show the perturbation equations for elastic solid crust regions. The ODEs are:

\begin{equation}
H_1' =
\left[
\frac{1}{2}(\lambda' - \nu') - \frac{\ell+1}{r}
\right] H_1
+ \frac{e^\lambda}{r}
\left[
H_2 + K - 16\pi(\varepsilon + p)V
\right]
\end{equation}
\begin{equation}
K' =
\frac{1}{r} H_2 + \frac{n+1}{r} H_1
+ \left(\frac{1}{2}\nu' - \frac{\ell+1}{r}\right) K
- \frac{8\pi(\varepsilon + p)e^{\lambda/2}}{r} W
\end{equation}
\begin{equation}
\begin{aligned}
H_0' &= K' - r e^{-\nu}\omega^2 H_1
- \left(\frac{1}{2}\nu' + \frac{\ell-1}{r}\right) H_0
- \left(\frac{1}{2}\nu' + \frac{1}{r}\right) H_2
+ \frac{\ell}{r} K
- \frac{16\pi}{r} T_2
\end{aligned}
\end{equation}
\begin{equation}
\begin{aligned}
W' &=
\frac{\ell+1}{r} W
+ r e^{\lambda/2}
\Biggl[
\frac{e^{-\nu}}{\Gamma p} X
- \frac{2(n+1)}{r^2} V
+ \frac{1}{2} H_2 + K
\Biggr]
\end{aligned}
\end{equation}
\begin{equation}
V' =
-\frac{1}{\mu r} T_2
+ \frac{e^{\lambda/2}}{r} W
+ \frac{2-\ell}{r} V
\end{equation}
\begin{equation}
\begin{aligned}
T_2' &=
-\frac{1}{2} r e^\lambda (\varepsilon + p) H_0
+ r e^{-\lambda/2}
\left(
X + \frac{1}{2r^2} e^{\nu/2} T_1
\right)
+ \left[
\frac{1}{2}(\lambda' - \nu') - \frac{\ell+1}{r}
\right] T_2 \\
&\quad
+ \left[
\frac{2 n \mu}{r}
- e^{\lambda-\nu} r\omega^2 (\varepsilon + p)
\right] V
+ e^{\lambda/2} p' W
\end{aligned}
\end{equation}

along with the algebraic equations:
\begin{equation}
H_2 = H_0 + 32\pi \mu V
\end{equation}
\begin{equation}
\begin{aligned}
\Biggl[
\frac{r e^{-\lambda}}{2}(r\nu' - 2) + (n+1)r
\Biggr] H_0
&=
r^2 e^{-\lambda}
\left[
\omega^2 r e^{-\nu} - \frac{n+1}{2}\nu'
\right] H_1
+ [
n_\ell r - \omega^2 r^3 e^{-\nu}- \frac{1}{4} r^2 e^{-\lambda}\nu'(r\nu' - 2)
] K \\
&\quad
+ 8\pi r^3 e^{-\nu/2} X
- 8\pi r T_1- 16\pi r e^{-\lambda} T_2
\end{aligned}
\end{equation}
\begin{equation}
\frac{1}{3} e^{-\nu/2} r^2 X
+ \frac{1}{4} \Gamma p\, T_1
=
\frac{\mu \Gamma p}{2}
\left[
2 e^{-\lambda/2} W - r^2 K + 2(n+1)V
\right]
\end{equation}

\subsection{Perturbation Equations for the Vaccum Regions}\label{sec:numerical}
The perturbation equations simplify exterior of the star. For the vaccum regions we have only 2 degrees of freedom. Here we refer \cite{lindblomQuadrupolEoScillationsNeutron1983,zerilliEffectivePotentialEvenParity1970,Miao:2025utd} and introduce the master variable $Z$. We have the relations between $H_1,K$ and $Z$ for the vaccum regions:
\begin{equation}
\begin{pmatrix}
\dfrac{nr^2 - 3nMr - 3M^2}{(r-2M)(nr+3M)} & r \\
\\
\dfrac{n(n+1)r^2 + 3nMr + 6M^2}{r^2(nr+3M)} & \dfrac{r-2M}{r}
\end{pmatrix}\begin{pmatrix}
Z  \\
\frac{dZ}{dr}
\end{pmatrix}=\\
\begin{pmatrix}
r^{\ell+1}H_1  \\
r^\ell K
\end{pmatrix}
\end{equation}
$Z$ obeys the master equation called Zerilli equation:
\begin{equation}
\frac{d^2 Z}{dr_*^2} = \left[ V_Z(r^*) - \omega^2 \right] Z
\end{equation}
Where $r_*=r+2M\ln{[r/(2M)-1]}$ and $V_Z$ is the effective potential:
\begin{equation}
\begin{aligned}
V_Z(r^*) &=
\frac{1 - \frac{2M}{r}}{r^3 (n r + 3M)^2}
[
2 n^2 (n+1) r^3
+ 6 n^2 M r^2\\
&+ 18 n M^2 r
+ 18 M^3
]   
\end{aligned}
\end{equation}.
In general we have the outgoing and ingoing solutions:
\begin{equation}
Z_{\rm{out}} =
e^{i\omega r_*}
\left[
1 + a_1 r^{-1} + a_2 r^{-2} + a_3 r^{-3}
+ \mathcal{O}(r^{-4})
\right]
\end{equation}
\begin{equation}
Z_{\rm{in}} =
e^{-i\omega r_*}
\left[
1 + a_1^* r^{-1} + a_2^* r^{-2} + a_3^* r^{-3}
+ \mathcal{O}(r^{-4})
\right]
\end{equation}
The coefficients are:
\begin{equation}
a_1 = \frac{i(n+1)}{\omega}
\end{equation}
\begin{equation}
a_2 =
- \frac{1}{2\omega^2}
\left[
n(n+1) + \frac{3i(n+2)}{n} M\omega
\right]
\end{equation}
\begin{equation}
\begin{aligned}
a_3 =
- \frac{i}{6\omega^3}
\Bigg[
& n(n-2)(n+1)
+ 3i(n+4)M\omega \\
& - \frac{18(2n+3)}{n^2} M^2 \omega^2
\Bigg]
\end{aligned}
\end{equation}
The general solution can be expressed as
\begin{equation}
Z = A_{\rm{out}} Z_{\rm{out}} + A_{\rm{in}} Z_{\rm{in}}.
\end{equation}

\section{Mode Excitation Problem with Near-Zone Boundary Conditions}
\label{app:andersson2025}
In this section we refer \cite{lindblom1997relativistic,Andersson:2025iyd} to have a brief introduction to the formalism which applies the near-zone boundary conditions.

Here we denote $\tilde{H}_0\equiv r^lH_0$. Near-zone boundary conditions indicate the behavior of $H_0$ at the surface could be interpreted as the response term and the tidal term:
\begin{equation}
\begin{aligned}
\tilde{H}_0(R,\omega)
=
A \left[1 + \alpha (R\omega)^2 \right] Q_2^2\!\left(\frac{R}{M} - 1\right)
\\+ B \left[1 + \beta (R\omega)^2 \right] P_2^2\!\left(\frac{R}{M} - 1\right)  
\end{aligned}
\end{equation}
\begin{equation}
\begin{aligned}
\tilde{H}_0'(R,\omega)
&=
A \Bigg\{
2\alpha \omega^2 R\, Q_2^2\!\left(\frac{R}{M} - 1\right)
+ \left[1 + \alpha (R\omega)^2 \right]
\left.\frac{d Q_2^2}{dr}\right|_{R/M-1}
\Bigg\} \\
&\quad
+ B \Bigg\{
2\beta \omega^2 R\, P_2^2\!\left(\frac{R}{M} - 1\right)
+ \left[1 + \beta (R\omega)^2 \right]
\left.\frac{d P_2^2}{dr}\right|_{R/M-1}
\Bigg\}
\end{aligned}
\end{equation}
\begin{equation}
P_2^2\!\left(\frac{r}{M} - 1\right)
=
3\left(\frac{r}{M}\right)^2
\left(1 - \frac{2M}{r}\right)
\end{equation}

\begin{equation}
\begin{aligned}
Q_2^2\!\left(\frac{r}{M} - 1\right)
&=
\left(\frac{r}{M}\right)^2
\left(1 - \frac{2M}{r}\right) \\
&\quad \times
\Biggl[
\frac{M(r - M)(2M^2 + 6Mr - 3r^2)}
{r^2 (r - 2M)^2}
+ \frac{3}{2} \ln\!\left(\frac{r}{r - 2M}\right)
\Biggr]
\end{aligned}
\end{equation}
Where $\alpha=-\frac{1}{2}$, $\beta=-\frac{11}{42}-\frac{107}{63}\frac{M}{r}$. $P^2_2,Q_2^2$ are the associated Legendre polynomials of the first and second kind of degree 2. We write the above relations as
\begin{equation}
\begin{aligned}
\tilde{H}_0 &= a_1 A + a_2 B, \\
\tilde{H}_0' &= a_3 A + a_4 B
\end{aligned}
\end{equation}
The tidal response could be expressed as:
\begin{equation}
    k_2=-\frac{4}{15}\left(\frac{M}{R}\right)^5\frac{A}{B}=\frac{4}{15}\left(\frac{M}{R}\right)^5\frac{a_4 \tilde{H}_0-a_2 \tilde{H}_0'}{a_3 \tilde{H}_0-a_1 \tilde{H}_0'}
\end{equation}
Each mode correspond a pole for $k_2$, namely
\begin{equation}
a_3 \tilde{H}_0 - a_1 \tilde{H}_0' = 0
\end{equation}
Thus $k_2$ could be further expressed as
\begin{equation}
k_\ell = -\frac{2\pi}{2\ell+1} \sum_n \frac{\mathcal{A}_n}{(\tilde{\omega} + \tilde{\sigma}_n - i\tilde{\gamma}_n)(\tilde{\omega} - \tilde{\sigma}_n - i\tilde{\gamma}_n)} \, .
\end{equation}
\begin{equation}
\mathcal{A}_n = -\frac{2\ell+1}{\pi} \tilde{\sigma}_n \lim_{\tilde{\omega} \to \tilde{\omega}_n} (\tilde{\omega} - \tilde{\omega}_n) k_l(\tilde{\omega})
= -\frac{2\ell+1}{\pi} \, \rm{Re}\,\tilde{\omega}_n \, \rm{Res}\, k_l(\tilde{\omega}_n) \, .
\label{eq:An}
\end{equation}
\begin{equation}
\rm{Res}\, k_\ell(\tilde{\omega}_n)
= -\frac{4}{15} \left(\frac{M}{R}\right)^5 
\left[ a_4 \tilde{H}_0 - a_2 \tilde{H}_0' \right]
\left[
\left. \frac{\partial (a_3 \tilde{H}_0 - a_1 \tilde{H}_0')}{\partial \tilde\omega} \right|_{\omega=\omega_n}
\right]^{-1} \, .
\end{equation}
Where the $\tilde{\sigma},\tilde{\gamma}$ denote real and imaginary parts, respectively. $\tilde{}$ denotes the frequency is normalized:
\begin{equation}
    \omega^2=\tilde{\omega}^2\frac{GM}{R^3}
\end{equation}
We consider $\sqrt{\mathcal{A}_n}$ as the so-called overlap integral in this formalism.

\section{Numerical Procedure}
\label{app:numerical} 
\subsection{Interior Solutions}
Our aim is to get the behaviour of the variables i.e. $H_0,H_1,K$ at the surface, thus we need to solve the interior perturbation equations. First we need to integrate from the center of the star. In practice we begin with a very small $r_c\approx1~\rm{mm}$. \cite{detweilerNonradialPulsationsGeneral1985,lindblomQuadrupolEoScillationsNeutron1983,Miao:2025utd} gives the conditions at the center:
\begin{equation}
    H_1(r_c)=\frac{2\ell K(r_c)+16\pi[\rho(r_c)+p(r_c)]W(r_c)}{\ell(\ell+1)}
\end{equation}
\begin{equation}
    V(r_c)=-\frac{1}{\ell}W(r_c)
\end{equation}
Once we specify $K(r_c),W(r_c)$ we have a complete initial conditions and then we have a solution.

At each interface, we should consider the junction conditions. At the fluid-fluid interface, we have:
\begin{equation}
    [H_1]_{r_i}=[K]_{r_i}=[W]_{r_i}=[X]_{r_i}=0.
\end{equation}
$[H_1]_{r_i}=H_1|_{r_i+\varepsilon}-H_1|_{r_i-\varepsilon}$ where $\varepsilon$ is very small. So $[H_1]_{r_i}=0$ means $H_1$ is continuous at the interface so are the other variables. 

At the fluid-solid interface we have:
\begin{equation}
    [H_0]_{r_i}=[H_1]_{r_i}=[K]_{r_i}=[W]_{r_i}=[T_2]_{r_i}=0.
\end{equation}
At the surface the Lagrangian perturbation of the pressure has to vanish, corresponding to:
\begin{equation}
    X(R)=0.
\end{equation}
In order to get the solutions which satisfy all these conditions, we should get a series of solutions from a series of independent initial conditions in each region and construct the solution by linear combinations. The combination coefficients are solved by these boundary conditions. Details are in \cite{lindblomQuadrupolEoScillationsNeutron1983,2015PhDT.......604K}.
\subsection{Find Modes with Outgoing Wave Boundary Condition}

We could get $H_1(R),K(R)$ from interior solutions. To find the real part of a mode, we proceed as follows:
\begin{enumerate}
    \item For each $\omega$, solve equations to get $H_1(R),K(R)$.
    \item Convert $H_1(R),K(R)$ to $Z(R),Z'(R)$ and integrate outwards. The terminal should be far enough, we choose $300R$. 
    \item Use the asymptotic behavior of Z (A21-A24) to solve $A_{out}(\omega),A_{in}(\omega)$.
    \item The mode frequency should satisfy $A_{in}(\omega)$. In practice, we search for $\omega$ that makes $A_{in}(\omega)$ a deep minimum.
\end{enumerate}
Iterative steps are proceed when we want to search the imaginary part:
\begin{enumerate}
    \item We get the initial approximate value of the real part above, say, $\omega_{r1}$.
    \item We search along the imaginary axis until we find $\omega_{i1}$ that makes $|\mathcal{R}(\omega=\omega_{r1}+i\omega_{i1})|$ reach a maximum.
    \item Then we fix the imaginary part and search along the real axis until we find $\omega_{r2}$ that makes $|\mathcal{R}(\omega=\omega_{r2}+i\omega_{i1})|$ reach a maximum.
    \item Iterate until the precision is achieved.
\end{enumerate}

\section{Corner Plot of Waveform Model Parameters}\label{app:corner}
\begin{figure}[H]
    \includegraphics[width=\textwidth]{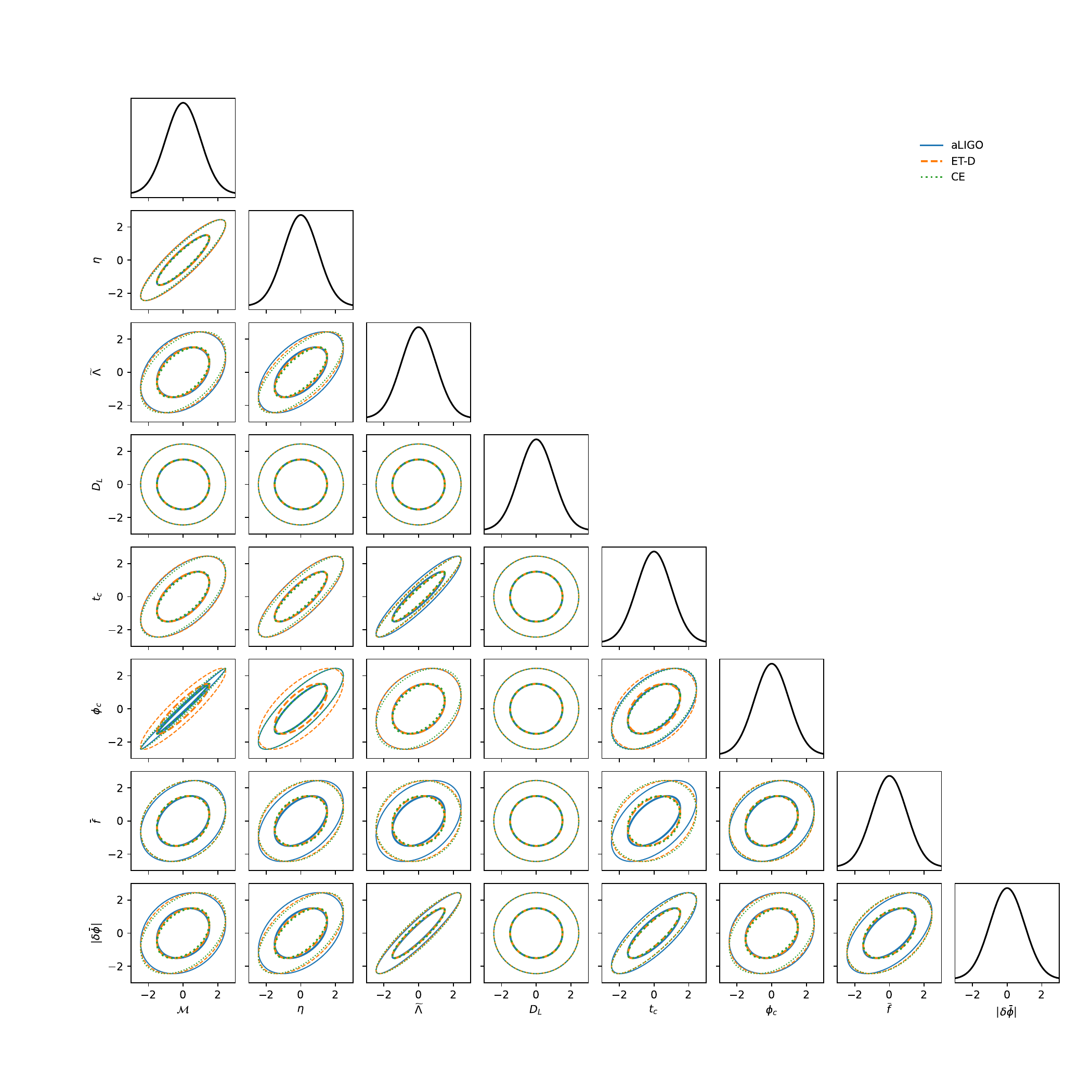}
    \caption{Fisher matrix forecast of parameter correlations for a $1.4+1.4~M_{\odot}$ binary at $100$ Mpc with an injected resonance frequency $\bar{f}=500$ Hz. For each detector, parameter offsets from their injected values are expressed in units of their marginalized standard deviations. Diagonal panels show the one-dimensional Gaussian marginals. In the off-diagonal panels, the inner and outer contours enclose $68\%$ and $95\%$ of the joint probability for each parameter pair after marginalization over the remaining parameters. For all three detectors, the uncertainty of the phase shift magnitude $|\delta\bar{{\phi}}|$  is  strongly correlated with that of the effective tidal deformability $\tilde{\Lambda}$, the coalescence time $t_c$ and the effective resonance frequency $\bar f$.}
    \label{fig:full parameter corner}
\end{figure}

Figure~\ref{fig:full parameter corner} visualizes the parameter covariance used in the Fisher forecast for a representative resonance at $\bar f=500\,\rm{Hz}$, with all offsets normalized by their marginalized standard deviations.
The tilted contours between $|\delta\bar{\phi}|$ and each of $\tilde{\Lambda}$, $t_c$, and $\bar f$ indicate strong correlations.
These degeneracies contribute to the increase in $\sigma_{\delta\bar{\phi}}$ relative to the idealized split-SNR estimate.

\twocolumngrid
\bibliographystyle{apsrev4-2}
\bibliography{refs}

\end{document}